# The Qutrit Bloch Sphere

**Vinod K. Mishra**

*313 Loganwood Ct, Joppa, MD 21085*

**Abstract**

It is very important to understand if a qutrit can be visualized in a 3-dimensional Bloch sphere. In this work, a mathematical model for performing this operation is presented.

## 1. Introduction

A qutrit density matrix is of order 3 and depends on 8 parameters. It lives in an 8-dimensional space which is impossible to visualize. A question arises if it is possible to do so in a 3-dimensional Bloch sphere. The earlier works [1-8] have tried to express Qutrit states as multi-dimensional object with dimensions more than 3 in general. It is hard to visualize those states. In the current approach, we have used the invariant properties of the qutrit states to present a scheme of visualization in 3 dimensions.

## 2. A general Qutrit density matrix

As is well known, the qutrit density matrix has 2 popular representations: (i) SU(3) using Gellman $\Lambda$-matrices, and (ii) Polarization using Spin-1 angular momentum matrices. The basic idea for visualizing a qutrit state is applicable to both, but here we will focus on the polarization representation represented by the following 3X3 Hermitian matrix $\hat{\rho}$.

$$\hat{\rho} = \frac{1}{3} 1_3 + \hat{\tau}$$

Here $1_3$ is a 3X3 unit matrix and in terms of polarization variables, $\hat{\tau}$ is given as

$$\hat{\tau} = \begin{bmatrix} \dfrac{y}{\sqrt{6}} + \dfrac{x}{\sqrt{2}} & -\left(\dfrac{a+\alpha_1}{\sqrt{2}} + i\dfrac{b+\beta_1}{\sqrt{2}}\right) & \alpha_2 + i\beta_2 \\ -\left(\dfrac{a+\alpha_1}{\sqrt{2}} - i\dfrac{b+\beta_1}{\sqrt{2}}\right) & -2\dfrac{y}{\sqrt{6}} & -\left(\dfrac{a-\alpha_1}{\sqrt{2}} + i\dfrac{b-\beta_1}{\sqrt{2}}\right) \\ \alpha_2 - i\beta_2 & -\left(\dfrac{a-\alpha_1}{\sqrt{2}} - i\dfrac{b-\beta_1}{\sqrt{2}}\right) & \dfrac{y}{\sqrt{6}} - \dfrac{x}{\sqrt{2}} \end{bmatrix}$$

The matrix $\hat{\rho}$ has unit trace ($Tr(\hat{\rho})=1$) and so:

$$Tr(\hat{\tau})=0.$$

The entries of $\hat{\tau}$ matrix are polarization variables and are represented by new symbols as given in Table 1 below.

**Table 1: Symbols for the matrix elements of $\hat{\tau}$**

| Symbol | $\gamma_1$ | $\gamma_2$ | $m_1$ | $n_1$ | $m_2$ | $n_2$ |
|---|---|---|---|---|---|---|
| Expression | $\dfrac{y}{\sqrt{6}} + \dfrac{x}{\sqrt{2}}$ | $\dfrac{y}{\sqrt{6}} - \dfrac{x}{\sqrt{2}}$ | $\dfrac{a+\alpha_1}{\sqrt{2}}$ | $\dfrac{b+\beta_1}{\sqrt{2}}$ | $\dfrac{a-\alpha_1}{\sqrt{2}}$ | $\dfrac{b-\beta_1}{\sqrt{2}}$ |

With redefinition of matrix elements, we have:

$$\hat{\tau} = \begin{bmatrix} \gamma_1 & -(m_1 + in_1) & \alpha_2 + i\beta_2 \\ -(m_1 - in_1) & -2\dfrac{y}{\sqrt{6}} & -(m_2 + in_2) \\ \alpha_2 - i\beta_2 & -(m_2 - in_2) & \gamma_2 \end{bmatrix}$$

## 3. The qutrit Bloch vectors

The trace inequalities [] applied to qutrits lead to the new Bloch-like vectors. Their derivation is given below.

### 3.1   1$^{st}$ Bloch vector and related expressions

*Matrix elements of $\hat{\rho}^2$*

The trace of $\hat{\rho}^2$ matrix is bound by an inequality given as:

$$Tr(\hat{\rho}^2)=Tr\left(\frac{1}{9}1_3+\frac{2}{3}\hat{\tau}+\hat{\tau}^2\right)\leq 1,$$

This leads to:

$$Tr(\hat{\rho}^2)=Tr\left(\frac{1}{9}1_3+\frac{2}{3}\hat{\tau}+\hat{\tau}^2\right)=\frac{1}{3}+Tr(\hat{\tau}^2)\leq 1$$

The relation $Tr(\hat{\tau})=0$ was used here. The elements of $\hat{\tau}^2$ matrix are given in terms of the new symbols as:

$$\hat{\tau}^2=\begin{bmatrix} \Gamma_1 & X_1+iY_1 & X_2+iY_2 \\ X_1-iY_1 & \Gamma_2 & X_3+iY_3 \\ X_2-iY_2 & X_3-iY_3 & \Gamma_3 \end{bmatrix}$$

Table 2 gives the $\hat{\tau}^2$ matrix elements in terms of the symbols of the $\hat{\tau}$ matrix.

**Table 2: $\hat{\tau}^2$ matrix elements in terms of $\hat{\tau}$ matrix elements**

| $\Gamma_1=\gamma_1^2+m_1^2+n_1^2+\alpha_2^2+\beta_2^2$ | $\Gamma_2=(\gamma_1+\gamma_2)^2+m_1^2+n_1^2+m_2^2+n_2^2$ | $\Gamma_3=\gamma_2^2+m_2^2+n_2^2+\alpha_2^2+\beta_2^2$ |
|---|---|---|
| $X_1=\gamma_2 m_1-(\alpha_2 m_2+\beta_2 n_2)$ | $X_2=(\gamma_1+\gamma_2)\alpha_2+m_1 m_2-n_1 n_2$ | $X_3=\gamma_1 m_2-\alpha_2 m_1-\beta_2 n_1$ |
| $Y_1=\gamma_2 n_1+\alpha_2 n_2-\beta_2 m_2$ | $Y_2=(\gamma_1+\gamma_2)\beta_2+m_2 n_1+m_1 n_2$ | $Y_3=\gamma_1 n_2+\alpha_2 n_1-\beta_2 m_1$ |

*Inequality relation and Bloch vector*

The $Tr(\hat{\rho}^2)=\frac{1}{3}+Tr(\hat{\tau}^2)\leq 1$ can be rewritten as:

$$\frac{3}{2}Tr(\hat{\tau}^2)=\frac{3}{2}(\Gamma_1+\Gamma_2+\Gamma_3)\leq 1,$$

In terms of the original parameters, we get:

$$\frac{3}{2}Tr(\hat{\tau}^2)=3\left[\frac{1}{2}(x^2+y^2)+A^2+B^2\right]\leq 1$$

Now the diagonal terms are related to the squares of Bloch vector's components leading to:

$$\frac{3}{2} Tr(\hat{\tau}^2) = u_1^2 + u_2^2 + u_3^2 = u^2 \leq 1$$

This Bloch vector is confined to a sphere with unit radius and has 3 degrees of freedom. Table 3 defines the Bloch vector inside a unit Bloch ball.

**Table 3: 1st Bloch vector components based on** $\frac{3}{2} Tr(\hat{\tau}^2) \leq 1$

| $u_1^2 = \frac{3}{2}\Gamma_1 = \frac{3}{2}(f_1^2 + f_2^2)$ | $u_2^2 = \frac{3}{2}\Gamma_2 = \frac{3}{2}\left(\frac{2}{3}y^2 + A^2\right) = y^2 + \frac{3}{2}A^2$ | $u_3^2 = \frac{3}{2}\Gamma_3 = \frac{3}{2}(f_1^2 - f_2^2)$ |
|---|---|---|
| $f_1^2 = \frac{x^2}{2} + \frac{y^2}{6} + \frac{A^2}{2} + B^2, \quad f_2^2 = \frac{xy}{\sqrt{3}} + C^2, \quad A^2 = a^2 + b^2 + \alpha_1^2 + \beta_1^2, \quad B^2 = \alpha_2^2 + \beta_2^2, \quad C^2 = a\alpha_1 + b\beta_1$ |||

### 3.2  2nd Bloch vector and related expressions

*Matrix elements of $\hat{\rho}^3$*

The cubed density matrix gives:

$$\hat{\rho}^3 = \frac{1}{27} 1_3 + \frac{1}{3}\hat{\tau} + \hat{\tau}^2 + \hat{\tau}^3.$$

With $\hat{\tau}^3$ as:

$$\hat{\tau}^3 = \begin{bmatrix} \Delta_{1R} + i\Delta_{1I} & \cdots & \cdots \\ \cdots & \Delta_{2R} + i\Delta_{2I} & \cdots \\ \cdots & \cdots & \Delta_{3R} + i\Delta_{3I} \end{bmatrix}$$

The subscripts $R$ and $I$ denote real and imaginary parts respectively. Off-diagonal terms of $\hat{\tau}^3$ have not been shown as they are not needed. It can be shown that $Tr(\hat{\tau}^3)$ has no imaginary part, i.e., $\Delta_{1I} + \Delta_{2I} + \Delta_{3I} = 0$. This leads to:

$$Tr(\hat{\tau}^3) = \Delta_{1R} + \Delta_{2R} + \Delta_{3R}$$

So, the real parts of the diagonal elements of $Tr(\hat{\tau}^3)$ are sufficient for further calculations.

*Inequality relation and Bloch vector*

The 2$^{nd}$ inequality depends both on $\hat{\rho}^2$ and $\hat{\rho}^3$ and is given by:

$$3\,Tr(\hat{\rho}^2) - 2\,Tr(\hat{\rho}^3) \leq 1$$

Which becomes:

$$Tr\left(6\hat{\tau} + \frac{9}{2}\hat{\tau}^2 - 9\hat{\tau}^3\right) = 9\,Tr\left(\frac{1}{2}\hat{\tau}^2 - \hat{\tau}^3\right) = 9\,Tr\begin{bmatrix} \frac{1}{2}\Gamma_1 - \Delta_{1R} & \cdots & \cdots \\ \cdots & \frac{1}{2}\Gamma_2 - \Delta_{2R} & \cdots \\ \cdots & \cdots & \frac{1}{2}\Gamma_3 - \Delta_{3R} \end{bmatrix} \leq 1$$

The above can be used to define a Bloch vector inside a unit Bloch ball leading to the squares of Bloch vector's vector components as:

$$v_1^2 = 9\left(\frac{1}{2}\Gamma_1 - \Delta_{1R}\right),\; v_2^2 = 9\left(\frac{1}{2}\Gamma_2 - \Delta_{2R}\right),\; v_3^2 = 9\left(\frac{1}{2}\Gamma_3 - \Delta_{3R}\right)$$

Then one can write:

$$v_1^2 + v_2^2 + v_3^2 = 9\,Tr\left(\frac{1}{2}\hat{\tau}^2 - \hat{\tau}^3\right) \leq 1$$

With length of the Bloch vector satisfying:

$$v = \sqrt{v_1^2 + v_2^2 + v_3^2} \leq 1$$

Table 4 below gives the expressions for the components of the Bloch vector based on the 2$^{nd}$ trace inequality as described in this section.

**Table 4: Real diagonal matrix elements of $\hat{\tau}^3$ and connected vector**

| Function | Definition | Final expression |
|---|---|---|
| $v_1^2 = 9\left(\frac{1}{2}\Gamma_1 - \Delta_{1R}\right)$ | $9\left[\left(\frac{1}{2} - \gamma_1\right)\Gamma_1 + (m_1 X_1 + n_1 Y_1) - (\alpha_2 X_2 + \beta_2 Y_2)\right]$ | $9(F_1 + F_3)$ |
| $v_2^2 = 9\left(\frac{1}{2}\Gamma_2 - \Delta_{2R}\right)$ | $9\left[\left(\frac{1}{2}+\gamma_1+\gamma_2\right)\Gamma_2 + (m_1 X_1 + n_1 Y_1) + (m_2 X_3 + n_2 Y_3)\right]$ | $9F_2$ |
| $v_3^2 = 9\left(\frac{1}{2}\Gamma_3 - \Delta_{3R}\right)$ | $9\left[\left(\frac{1}{2}-\gamma_2\right)\Gamma_3 + (m_2 X_3 + n_2 Y_3) - (\alpha_2 X_2 + \beta_2 Y_2)\right]$ | $9(F_1 - F_3)$ |

$$F_1 = \frac{3x^2 + y^2}{12} - \frac{y(9x^2 + y^2)}{6\sqrt{6}} + \frac{A^2}{4} + \frac{(1-\sqrt{6}\,y)B^2}{2} - \sqrt{2}\,xC^2 - D^3$$

$$F_2 = \frac{1}{3}y^2\left(1 + 2\sqrt{\frac{2}{3}}\,y\right) + \frac{1}{2}A^2(1 + \sqrt{6}\,y) - \sqrt{2}\,xC^2 - D^3$$

$$F_3 = \frac{x}{2\sqrt{2}}\left\{\sqrt{\frac{2}{3}}\,y - (x^2 + y^2)\right\} - \frac{x}{\sqrt{2}}(A^2 + B^2) + \frac{1}{2}C^2$$

$$A^2 = a^2 + b^2 + \alpha_1^2 + \beta_1^2, \quad B^2 = \alpha_2^2 + \beta_2^2, \quad C^2 = a\alpha_1 + b\beta_1,$$

$$D^3 = \alpha_2(a^2 - b^2 - \alpha_1^2 + \beta_1^2) + 2\beta_2(ab - \alpha_1\beta_1)$$

This leads to:

$$9\text{Tr}\left[\frac{1}{2}\hat{\tau}^2 - \hat{\tau}^3\right] = 9\left[\frac{x^2 + y^2}{2} + \frac{y(y^2 - 3x^2)}{\sqrt{6}} + \left(1 + \sqrt{\frac{3}{2}}\,y\right)A^2 + (1 - \sqrt{6}\,y)B^2 - 3\sqrt{2}\,xC^2 - 3D^3\right] \leq 1$$

### 3.3   3rd Bloch vector and related expressions

The density matrix obeys $Tr\,\hat{\rho} = 1$, and is given as:

$$\hat{\rho} = \frac{1}{3}\mathbf{1}_3 + \hat{\tau} = \begin{bmatrix} \frac{1}{3}+\gamma_1 & -(m_1+in_1) & \alpha_2+i\beta_2 \\ -(m_1-in_1) & \frac{1}{3}-2\frac{y}{\sqrt{6}} & -(m_2+in_2) \\ \alpha_2-i\beta_2 & -(m_2-in_2) & \frac{1}{3}+\gamma_2 \end{bmatrix}$$

The qutrit has 8 degrees of freedom out of which 6 have given us the two vectors **u** and **v** discussed earlier. The third vector **w** based on the above can be written as:

$$w_1^2 = \frac{1}{3} + \gamma_1, \quad w_2^2 = \frac{1}{3} - (\gamma_1 + \gamma_2), \quad w_3^2 = \frac{1}{3} + \gamma_2$$

It obeys

$$w_1^2 + w_2^2 + w_3^2 = 1$$

It is a unit vector and so has only two degrees of freedom. Together with the other two given earlier, we have 8 degrees of freedom, which can represent qutrit completely inside a Bloch sphere without needing an 8-dimensional space.

## 4. EXAMPLES

**Case I**: Only two variables (denoted by $s, t$) are nonzero

This case has been discussed in Ref [8] and resulting 2-dimensional plots have been given. The Table 5.1 gives the 1$^{st}$ inequality cases in two variables for both Ref [8] and current approach.

**Table 5: 1$^{st}$ inequality with two variables ($s, t$)**

| Clusters | 1$^{st}$ inequality: (1) Ref [8] and (2) this work | Associated vector components |
|---|---|---|
| I - $(x, y)$ | (1) $s^2 + t^2 \leq \frac{2}{3}$ (2) $\frac{3}{2}(s^2 + t^2) \leq 1$ | $u_1^2 = \frac{3}{4}s^2 + \frac{1}{4}t^2 + \frac{\sqrt{3}}{2}st$ <br> $u_2^2 = t^2$ <br> $u_3^2 = \frac{3}{4}s^2 + \frac{1}{4}t^2 - \frac{\sqrt{3}}{2}st$ |
| II - $(y, \alpha_2), (y, \beta_2)$ | (1) $s^2 + 2t^2 \leq \frac{2}{3}$ (2) $\frac{3}{2}(s^2 + 2t^2) \leq 1$ | $u_1^2 = \frac{1}{4}s^2 + \frac{3}{2}t^2$, $u_2^2 = s^2$ <br> $u_3^2 = \frac{1}{4}s^2 - \frac{3}{2}t^2$ |
| III - $(a, \alpha_2), (\beta_1, \alpha_2)$ | (1) $2(s^2 + t^2) \leq \frac{2}{3}$ | $u_1^2 = \frac{3}{4}s^2 + \frac{3}{2}t^2$, $u_2^2 = \frac{3}{2}s^2$ |

| | | (2) $3(s^2+t^2) \leq 1$ | $u_3^2 = \frac{3}{4}s^2 + \frac{3}{2}t^2$ |
|---|---|---|---|
| IV - | $(b,\alpha_2),(\alpha_1,\alpha_2)$ | Same as cluster III | Same as cluster III |
| V - | $(y,a),(y,b),$ $(y,\alpha_1),(y,\beta_1)$ | Same as cluster II | Same as cluster II |
| VI - | $(a,b),(a,\alpha_1),(a,\beta_1),$ $(a,\beta_2),(b,\alpha_1),(b,\beta_1),(b,\beta_2),$ $(\alpha_1,\beta_1),\ (\alpha_1,\beta_2),\ (\beta_1,\beta_2),$ $(\alpha_2,\beta_2)$ | (1) $2(s^2+t^2) \leq \frac{2}{3}$ (2) $3(s^2+t^2) \leq 1$ | $u_1^2 = \frac{3}{4}s^2 + \frac{3}{4}t^2$ $u_2^2 = \frac{3}{2}s^2 + \frac{3}{2}t^2$ $u_3^2 = \frac{3}{4}s^2 + \frac{3}{4}t^2$ |
| VII - | $(x,a),(x,b),(x,\alpha_1),$ $(x,\beta_1),(x,\alpha_2),(x,\beta_2)$ | Same as cluster II | Same as cluster II |

The Table 5.2a gives the 2$^{nd}$ inequality cases for two variables (except for cluster VI).

**Table 5.2a: 2$^{nd}$ inequality with two variables ($s$, $t$)**

| Clusters | 2$^{nd}$ inequality: (1) Ref [8] and (2) this work | associated vector components |
|---|---|---|
| I - $(x,y)$ | (1) $\frac{1}{9} - \frac{1}{2}(s^2+t^2)\frac{+t}{\sqrt{6}}(3s^2-t^2) \geq 0$ (2) $\frac{9}{2}(s^2+t^2) - \frac{3\sqrt{3}}{\sqrt{2}}(3s^2-t^2)t \leq 1$ | $v_1^2 = 9(F_1+F_3),\ v_2^2 = 9F_2$ $v_3^2 = 9(F_1-F_3)$ $F_1 = \frac{3s^2+t^2}{12} - \frac{t(9s^2+t^2)}{6\sqrt{6}}$ $F_2 = \frac{t^2}{3}\left(1+2\sqrt{\frac{2}{3}}t\right)$ $F_3 = \frac{s}{2\sqrt{2}}\left\{\sqrt{\frac{2}{3}}t - (s^2+t^2)\right\}$ |
| II - $(y,\alpha_2),(y,\beta_2)$ | (1) $\frac{1}{9} - \left(\frac{s^2}{2}+t^2\right)\frac{+s}{\sqrt{6}}(6t^2-s^2) \geq 0$ | $v_1^2 = 9F_1 = v_3^2,\ v_2^2 = 9F_2$ |
| | (2) $9\left(\frac{s^2}{2}+t^2\right)-\frac{3\sqrt{3}}{\sqrt{2}}(6t^2-s^2)s\leq 1$ | $F_1=\frac{s^2}{12}-\frac{s^3}{6\sqrt{6}}+\frac{(1-\sqrt{6}s)t^2}{2}$ $F_2=\frac{s^2}{3}\left(1+2\sqrt{\frac{2}{3}}s\right), F_3=0$ |
| III - $(a,\alpha_2),(\beta_1,\alpha_2)$ | (1) $\frac{1}{9}-(s^2+t^2)+3s^2t\geq 0$ (2) $9(s^2+t^2)-27s^2t\leq 1$ | $v_1^2=9F_1=v_3^2, v_2^2=9F_2$ $F_1=\frac{s^2+2t^2}{4}-s^2t$ $F_2=-s^2t, F_3=0$ |
| IV - $(b,\alpha_2),(\alpha_1,\alpha_2)$ | (1) $\frac{1}{9}-(s^2+t^2)-3s^2t\geq 0$ (2) $9(s^2+t^2)+27s^2t\leq 1$ | $v_1^2=9F_1=v_3^2, v_2^2=9F_2$ $F_1=\frac{s^2}{4}+t^2\left(\frac{1}{2}+s^2\right)$ $F_2=s^2\left(\frac{1}{2}+t\right), F_3=0$ |
| V - $(y,a),(y,b),$ $(y,\alpha_1),(y,\beta_1)$ | (1) $\frac{1}{9}-\left(\frac{s^2}{2}+t^2\right)-\frac{s}{\sqrt{6}}(3t^2+s^2)\geq 0$ (2) $9\left(\frac{s^2}{2}+t^2\right)+\frac{3\sqrt{3}}{\sqrt{2}}(3t^2+s^2)s\leq 1$ | $v_1^2=9F_1=v_3^2, v_2^2=9F_2$. $F_1=\frac{s^2}{12}-\frac{s^3}{6\sqrt{6}}+\frac{t^2}{4}, F_3=0$ $F_2=\frac{s^2}{3}\left(1+2\sqrt{\frac{2}{3}}s\right)\frac{+(1+\sqrt{6}s)t^2}{2}$ |
| VI It is given below in a separate table as it has many different cases. | | |
| VII - $(x,a),(x,b),(x,\alpha_1),$ $(x,\beta_1),(x,\alpha_2),(x,\beta_2)$ | (1) $\frac{1}{9}-\left(\frac{s^2}{2}+t^2\right)\geq 0$ (2) $9\left(\frac{s^2}{2}+t^2\right)\leq 1$ | $v_1^2=9(F_1+F_3), v_2^2=9F_2$ $v_3^2=9(F_1-F_3)$ $F_1=\frac{s^2+t^2}{4}, F_2=\frac{t^2}{2}, F_3=-\frac{st^2}{\sqrt{2}}$ |

The Table 5.2b gives the 2nd inequality cases in two variables for cluster VI.

**Table 5.2b: 2$^{nd}$ inequality with two variables ($s$, $t$): Cluster VI**

2$^{nd}$ inequality: **(1) Ref [8]**: $\frac{1}{9} - (s^2 + t^2) \geq 0$  **(2) this work** $9(s^2 + t^2) \leq 1$

| Clusters | associated vector components |
|---|---|
| $(a,b), (a,\beta_1), (b,\alpha_1)$ | $v_1^2 = 9F_1 = v_3^2,\ v_2^2 = 9F_2,\ F_1 = \frac{s^2+t^2}{4},\ F_2 = \frac{s^2+t^2}{2},\ F_3 = 0$ |
| $(a,\alpha_1), (b,\beta_1)$ | $v_1^2 = 9(F_1 + F_3),\ v_2^2 = 9F_2,\ v_3^2 = 9(F_1 - F_3)$ <br> $F_1 = \frac{s^2+t^2}{4},\ F_2 = \frac{s^2+t^2}{2},\ F_3 = \frac{st}{2}$ |
| $(a,\beta_2), (b,\beta_2),\ \alpha_1,\beta_2$ | $v_1^2 = 9F_1 = v_3^2,\ v_2^2 = 9F_2,\ F_1 = \frac{s^2+2t^2}{4},\ F_2 = \frac{s^2}{2},\ F_3 = 0$ |
| $(\alpha_1,\beta_1)$ | $v_1^2 = 9F_1 = v_3^2,\ v_2^2 = 9F_2,\ F_1 = \frac{s^2+t^2}{4},\ F_2 = \frac{s^2+t^2}{2},\ F_3 = 0$ |
| $(\alpha_2,\beta_2)$ | $v_1^2 = 9F_1 = v_3^2,\ v_2^2 = 0,\ F_1 = \frac{s^2+t^2}{2},\ F_2 = 0,\ F_3 = 0$ |
| $(\beta_1,\beta_2)$ | $v_1^2 = 9F_1 = v_3^2,\ v_2^2 = 9F_2,\ F_1 = \frac{s^2+2t^2}{4},\ F_2 = \frac{s^2}{2},\ F_3 = 0$ |

**Case II**: The case when four variables (denoted by $s, t, u, v$)) are nonzero

This case is given partially for the sake of illustration showing the utility of the current approach for the qutrit state visualization. This shows also the inability of earlier approach to perform this task.

**Table 5.3: 1$^{st}$ inequality with four variables ($s$, $t$, $u$, $v$)**

| $(s,t,u,v)$ | 1$^{st}$ inequality |
|---|---|
| $(x,y,\alpha_1,\beta_1)$ | $v_1^2 = 9(F_1 + F_3),\ v_2^2 = 9F_2,\ v_3^2 = 9(F_1 - F_3)$ <br> $F_1 = \frac{3s^2+t^2}{12} - \frac{t(9s^2+t^2)}{6\sqrt{6}} + \frac{u^2+v^2}{4}$ <br> $F_2 = \frac{t^2}{3}\left(1 + 2\sqrt{\frac{2}{3}}t\right) + \frac{u^2+v^2}{2}(1+\sqrt{6}t)$ |
| | $F_3 = \dfrac{s}{2\sqrt{2}}\left\{\sqrt{\dfrac{2}{3}}t - (s^2+t^2) - 2(u^2+v^2)\right\}$ |
| $(a, b, \alpha_2, \beta_2)$ | $v_1^2 = 9F_1 = v_3^2,\ v_2^2 = 9F_2$ |
| | $F_1 = \dfrac{s^2+t^2}{4} + \dfrac{u^2+v^2}{2} - (s^2-t^2)u - 2stv$ |
| | $F_2 = \dfrac{s^2+t^2}{2} - (s^2-t^2)u - 2stv,\ F_3 = 0$ |
| $(a, b, \alpha_1, \beta_1)$ | $v_1^2 = 9(F_1+F_3),\ v_2^2 = 9F_2,\ v_3^2 = 9(F_1-F_3)$ |
| | $F_1 = \dfrac{A^2}{4},\ F_2 = \dfrac{A^2}{2},\ F_3 = \dfrac{C^2}{2}$ |

The earlier methods need more than three dimensions so are unable to help with visualization. In the new method, the dimensions are captured by the vector components and are thus able to perform that task.

## 5. CONCLUSIONS

In this work we have shown how to extract a qutrit state from a 4 X 4 general 2-qubit state matrix. The qutrit matrix element expressions have been given in terms of original 2-qubit matrix elements. In future work, we will explore other aspects of this transformation applicable to quantum computing and communication.

## 6. ACKNOWLEDGEMENT

I thank Prof. Warner Miller (Florida Atlantic University) for encouragement and discussions.